\documentclass[%
 aip,
 amsmath,amssymb,
preprint,%
]{revtex4-1}
\usepackage{graphicx}
\usepackage{dcolumn}
\usepackage{bm}

\usepackage[utf8]{inputenc}
\usepackage[T1]{fontenc}
\usepackage{mathptmx}
\usepackage{etoolbox}
\usepackage{xcolor}
\usepackage{soul}
\usepackage{ulem}

\makeatletter
\def\@email#1#2{%
 \endgroup
 \patchcmd{\titleblock@produce}
  {\frontmatter@RRAPformat}
  {\frontmatter@RRAPformat{\produce@RRAP{*#1\href{mailto:#2}{#2}}}\frontmatter@RRAPformat}
  {}{}
}%
\makeatother
\begin{document}
\title[]{Strain-induced magnetic damping anomaly in La$_{1-x}$Sr$_{x}$MnO$_{3}$ ($x$~=~$0.3$ -- $0.5$) thin films}
\author{Ryotaro Arakawa}
\author{Sachio Komori*}
\author{Tomoyasu Taniyama*}
\affiliation{ 
Department of Physics, Nagoya University, Nagoya 464-8602, Japan
}
\email[]{taniyama.tomo@nagoya-u.jp}
\email[Author to whom correspondence should be addressed:]{komori.sachio.h0@f.mail.nagoya-u.ac.jp}

\date{\today}
\begin{abstract}
Magnetic properties of La$_{1-x}$Sr$_x$MnO$_{3}$ (LSMO) are highly sensitive to various factors such as the Sr doping level $x$, lattice strain, and oxygen stoichiometry due to the strongly correlated nature of 3$d$ electrons. For the development of energy-efficient spintronic devices with ultra-low magnetic damping of LSMO, a thorough understanding of its complex magnetization dynamics is of great importance. In this work, we have measured ferromagnetic resonance of LSMO thin films on Nb-doped SrTiO$_3$ (Nb-STO) substrates over a wide temperature and frequency range and observed an anomalous increase in the Gilbert damping constant and a decrease in the effective saturation magnetization at temperatures below 100 K. The anomalies become more pronounced as the LSMO thickness decreases while they are not observed for LSMO on (LaAlO$_3$)$_{0.3}$(Sr$_2$TaAlO$_6$)$_{0.7}$ substrates with relatively small epitaxial strain. The results suggest that the epitaxial strain-induced magnetically dead layer at the LSMO/Nb-STO interface acts as a spin sink and leads to the anomalies in the magnetization dynamics. 
\end{abstract}
\maketitle
Strongly correlated oxide systems exhibit complex and rich physics due to the strong coupling between spin, lattice and orbital degrees of freedom. A ferromagnetic manganite La$_{1-x}$Sr$_x$MnO$_3$ (LSMO) is a widely studied strongly correlated electron system with advantageously high spin-polarization close to 100\% at room temperature particularly around $x~\approx$~0.3. The half-metallic nature of LSMO leads to ultra-low Gilbert damping, making it a promising candidate for energy-efficient spintronic memory and spin-wave devices.\cite{Qin2017,Bowen2003} However, the key factors that determine the Gilbert damping of LSMO have not been fully understood. This is mainly because the magnetic properties of LSMO, particularly in thin films, are affected by several factors such as the doping level $x$, epitaxial strain, and film thickness. A temperature-dependent broadband ferromagnetic resonance (FMR) measurement is a powerful method to investigate the origins of magnetic damping. Recently, a strong temperature dependence of the Gilbert damping constant was reported for LSMO with $x$~=~0.3, which was explained by the Kambersky's torque correlation model in conjunction with an additional effect of spin pumping into an adjacent magnetically active dead layer.\cite{Haspot2022,Wang2023} To gain broader and deeper insight into the mechanisms of magnetic damping and to develop a thorough understanding of magnetization dynamics, FMR studies of LSMO with different doping levels and strain states are needed.

Here, we present a temperature-dependent broadband FMR study of LSMO ($x$~=~0.5) thin films near the ferromagnetic-antiferromagnetic phase boundary. We show an anomalous increase in the Gilbert damping constant ($\alpha$) and a decrease in the effective saturation magnetization ($M_\text{eff}$) below 100 K. From LSMO-thickness-dependent FMR measurements, we are able to correlate the anomalies in $\alpha$ and $M_\text{eff}$ to the magnetically dead layer localized at the LSMO/substrate interface, which serves as a spin sink and affects the magnetization dynamics. The results provide deeper insight into magnetic damping in LSMO and related transition metal oxides.

LSMO ($x$~=~0.37 and 0.5) thin films with a thickness range of 10 -- 38~nm were epitaxially grown on (001)-oriented Nb-doped (0.5 wt\%) SrTiO$_3$ (Nb-STO) and (LaAlO$_3$)$_{0.3}$(Sr$_2$TaAlO$_6$)$_{0.7}$ (LSAT) substrates by pulsed laser deposition (the fourth harmonic of a Q-switched Nd-YAG laser; wavelength $\lambda$~=~266 nm) with a laser fluence of 8.4 J/cm$^2$ and a laser frequency of 10 Hz at an oxygen pressure of 500 mTorr and a temperature of 750$^{\circ}$C. After the deposition, the samples were cooled down to room temperature in 2 hours under an oxygen pressure of 200 Torr. 

\begin{figure}[h]
    \includegraphics[width=8.5cm]{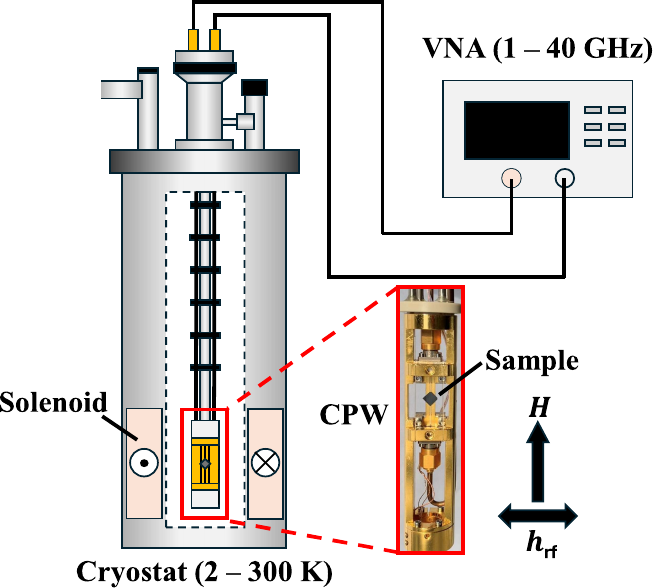}
    \caption{\label{fig:FMR_setup} Schematic of the VNA-FMR setup. The sample was placed on the CPW and the static magnetic field $\bm{H}$ and the rf magnetic field $\bm{h}_\text{rf}$ were applied orthogonally to each other, both within the sample plane.}
\end{figure}

Magnetization dynamics was investigated using a broadband FMR setup with a vector network analyzer (VNA) and a coplanar wave guide (CPW) (see Fig.~\ref{fig:FMR_setup}). 
The sample was covered with electron beam resist (ZEP-520A) for electrical insulation and attached face down on the CPW using high vacuum grease. The CPW was placed in a pulse-tube cryogen-free cryostat with a superconducting solenoid magnet and measurements were performed over a temperature range of 2 -- 300 K. A microwave signal passing through the CPW is partially absorbed by an excitation of the FMR, which is measured by the $S_{21}$ transmission parameter. The microwave power was set to +8 dBm and the frequency ($f$) was swept from 0.1 to 40 GHz at fixed external magnetic fields ($H$) from 0 to 15000~Oe applied along the in-plane [110] crystallographic orientation of LSMO (along the magnetic easy axis). The absorption spectrum at $H~$=$~0$ was subtracted from all the spectra recorded at $H~$>$~0$ as a background and the resulting data were denoted as $\Delta$Mag$S_{21}$.

\begin{figure}[h]
    \includegraphics[width=15cm]{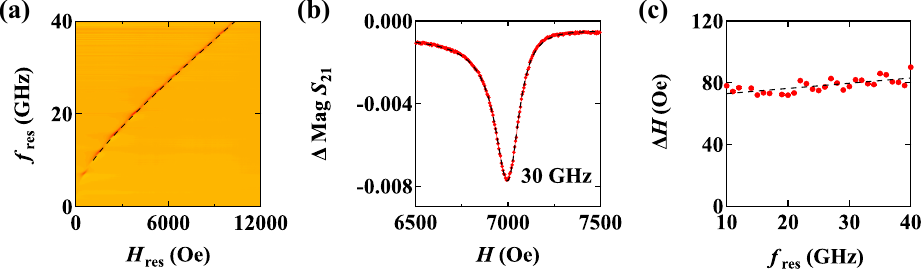}
    \caption{\label{fig:FMR} (a) Typical FMR absorption spectrum $\Delta$Mag$S_{21}$ for 28-nm-thick La$_{0.5}$Sr$_{0.5}$MnO$_3$ on Nb-STO at 100~K. (b) Absorption spectrum at 30 GHz. (c) Half width at half maximum value $\Delta H$ versus resonance frequency $f_\text{res}$. The dashed curves in (a), (b), and (c) represent least-squares line fits by Kittel's formula, Lorentzian functions, and the linear relationship between $\alpha$ and $f_\text{res}$, respectively.}
\end{figure}

Figure~\ref{fig:FMR}(a) shows a typical $\Delta$Mag$S_{21}$ absorption spectrum, from which the effective saturation magnetization ($M_{\text{eff}}$) and the in-plane anisotropy field ($K_1$) were obtained by fitting the resonance frequency ($f_{\text{res}}$) and the resonance magnetic field ($H_\text{res}$) to the Kittel's formula: $f_{\text{res}}~=~\gamma/2\pi\sqrt{{(H_{\text{res}} + 4\pi M_{\text{eff}} + 2K_1 / M_{\text{eff}})}(H_{\text{res}} + 2K_1 / M_{\text{eff}})}$, where $\gamma$ is the gyromagnetic ratio for an electron spin.\cite{Kittel1948} 
To estimate $\alpha$, $\Delta$Mag$S_{21}$ curves at different $f_{\text{res}}$ ranging from 10 to 40 GHz are extracted from the $\Delta$Mag$S_{21}$ absorption spectrum [see Fig.~\ref{fig:FMR}(b) for a typical $\Delta$Mag$S_{21}$ curve]. The $\Delta$Mag$S_{21}$ curves were fitted by a linear combination of symmetric and antisymmetric Lorentzian functions\cite{Kumar2017,Usami2021}:
    \begin{eqnarray}
    \Delta \text{Mag}S_{21}=
    C_{\text{sym}}\frac{(\Delta H)^2}{(\Delta H)^2 + (H-H_{\text{res}})^2}
    +C_{\text{asym}}\frac{\Delta H(H-H_{\text{res}})^2}{(\Delta H)^2 + (H-H_{\text{res}})^2} + B,
    \end{eqnarray}
where $C_{\text{sym}}$ and $C_{\text{asym}}$ are the amplitudes of the symmetric and antisymmetric Lorentzian functions, respectively, and $B$ is the constant term to take account for the influence of the background. The ratio of $C_{\text{sym}}$ and $C_{\text{asym}}$ relates to the phase shift between the microwave current and the FMR response. The fit gives an estimate of the half-width at half maximum ($\Delta H$) of $\Delta$Mag$S_{21}$. From the linear scaling of $\Delta H$ with $f_\text{res}$: $\Delta H$ $ =\Delta H_0 + 2\pi f_{\text{res}}\alpha/\gamma$ [see Fig.~\ref{fig:FMR}(c)], we obtained effective $\alpha$, where $H_0$ is the zero-frequency line broadening due to intrinsic magnetic inhomogeneity. \cite{Platow1998, Lee2016}

\begin{figure}[h]
    \includegraphics[width=13cm]{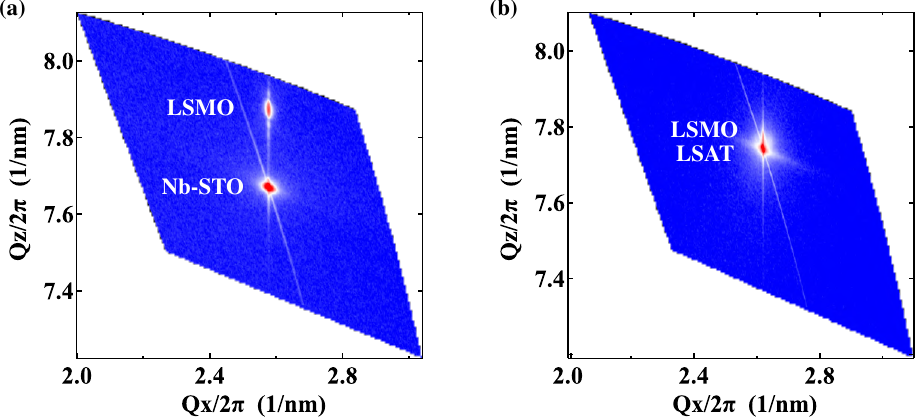}
    \caption{\label{fig:RSM} X-ray diffraction reciprocal space mapping around the (103) diffraction peak from 38-nm-thick La$_{0.5}$Sr$_{0.5}$MnO$_3$ on (a) Nb-STO and (b) LSAT.}
\end{figure}

Figures~\ref{fig:RSM}(a) and \ref{fig:RSM}(b) show X-ray diffraction reciprocal space mapping around the (103) peak of 38-nm-thick LSMO with $x$~=~0.5 on Nb-STO and LSAT, respectively, confirming coherent growth of LSMO without detectable lattice relaxation. 
From the peak position, the anisotropic lattice strain ($c/a$ ratio of the lattice constants) of LSMO on LSAT is estimated to be close to 1, while it is 0.982 on Nb-STO, indicating the presence of tensile strain along the in-plane direction on Nb-STO. The pseudo-cubic lattice parameter of bulk LSMO with $x$~=~0.5 (3.847~\AA) \cite{Spooren2005} is close to that of LSAT (3.868~\AA )\cite{Zhang2022}, but notably smaller than that of Nb-STO (3.903~\AA)\cite{Albargi2018}, hence a larger tensile strain is expected in LSMO/Nb-STO, consistent with the smaller $c/a$ ratio. The $c/a$ ratio tends to decrease down to 0.967 with decreasing thickness of LSMO on Nb-STO (see Supplementary Information for details).

Figure~\ref{fig:Damping_x} shows the temperature dependence of $\alpha$ for LSMO ($x$~=~0.37)/Nb-STO with $x$~=~0.37 giving the highest Curie temperature\cite{Tokura2006}
and $x$~=~0.5 near the ferromagnetic--antiferromagnetic phase boundary.\cite{Tokura2006}
The $\alpha$ of LSMO ($x$~=~0.37) does not show a strong temperature dependence down to 100~K except at 300~K (near the Curie temperature), while the $\alpha$ of LSMO ($x$~=~0.5) decreases with decreasing temperature down to 100 K. The strong temperature dependence above 100~K can be attributed to the large volume fraction of the paramagnetic phase in overdoped LSMO ($x$~=~0.5) with a relatively low Curie temperature. It has been reported that the ferromagnetic and paramagnetic phases coexist in LSMO near the antiferromagnetic and ferromagnetic phase boundary ($x$~=~0.55) near room temperature,\cite{Li2002} and therefore, a considerable amount of the paramagnetic phase can also exist in our LSMO ($x$~=~0.5). The paramagnetic phase can become a spin sink and its volume fraction changes with temperature, causing the strong temperature dependence of $\alpha$. LSMO ($x$~=~0.37) with a higher Curie temperature has a less amount of the paramagnetic phase, which explains the absence of the strong temperature dependence of $\alpha$. 

Below 100 K, $\alpha$ increases anomalously for both $x$~=~0.37 and 0.5. 
A similar increase in the magnetic damping at low temperatures has been recently reported in LSMO with $x$~=~0.3 and its potential relation with a magnetically dead layer has been pointed out.
A Sudden decrease in $\alpha$ below 10 K was also attributed the dead layer and was explained by its spin-glass-like transition that decreases spin relaxation channels below 10~K.\cite{Haspot2022} From here on, we focus on LSMO with $x$~=~0.5 showing a relatively large increase of $\alpha$ and discuss the origin of this damping anomaly including a role of the dead layer. 

In Figs.~\ref{fig:Damping_substrate}(a) and \ref{fig:Damping_substrate}(b), we plot the temperature and thickness dependence of $\alpha$ for LSMO ($x$~=~0.5) on Nb-STO and LSAT, respectively. 
For LSMO/Nb-STO, a remarkable enhancement of $\alpha$ below 100~K is observed for all the LSMO thicknesses. The enhancement is largest for the thinnest LSMO (10 nm), suggesting that the origin of the anomalous damping enhancement is located near the LSMO/Nb-STO interface. In contrast, the $\alpha$ of LSMO/LSAT shows a slight monotonic increase with decreasing temperature over the entire temperature range. To gain further insight into the $\alpha$ enhancement at low temperatures, we compare the effective saturation magnetization $M_{\text{eff}}$ estimated from the FMR absorption spectra and the DC saturation magnetization $M_\text{s}$ measured by a superconducting quantum interference device (SQUID) at $H$~=~500 Oe. In Figs.~\ref{fig:Manetization}(a) and~\ref{fig:Manetization}(b), we show the temperature dependence of $M_{\text{eff}}$ and $M_\text{s}$ for the corresponding samples in Figs. \ref{fig:Damping_substrate}(a) and \ref{fig:Damping_substrate}(b), respectively. Note that FMR directly probes the saturation magnetization of a ferromagnetic phase while the DC saturation magnetization measured by SQUID is the average of the magnetic moment of both ferromagnetic and paramagnetic phases in LSMO, meaning that the coexistence of ferromagnetic and paramagnetic phases in LSMO leads to the discrepancy between $M_{\text{eff}}$ and $M_\text{s}$. For LSMO/Nb-STO, $M_\text{s}$ is remarkably smaller than $M_{\text{eff}}$, which is in sharp contrast to LSMO/LSAT showing the comparable $M_\text{s}$ and $M_{\text{eff}}$ values over the whole temperature range. The results suggest the presence of a notable amount of paramagnetic phase in LSMO/Nb-STO, while it is absent in LSMO/LSAT.

The magnetic ground state of LSMO is highly sensitive to epitaxial strain particularly near the FM-AFM phase boundary ($x \approx$ 0.5). According to the magnetic phase diagram in Ref.~\cite{Konishi1999,Gutierrez2014}, LSMO ($x$~=~0.5)/LSAT with $c/a\approx$ 1 has a ferromagnetic ground state while LSMO ($x$~=~0.5)/STO with $c/a$~<~1 has an A-type antiferromagnetic ground state. This is in good agreement with our results: LSMO/LSAT with $c/a\approx$~1 shows larger $M_\text{s}$ than LSMO/Nb-STO ($c/a$~=~0.969 -- 0.982).

Figure~\ref{fig:Manetization}(c) shows the LSMO-thickness dependence of $M_\text{s}$ per unit area at 2 and 275~K for LSMO/Nb-STO. 
Extrapolation from the linear fit reveals the presence of a magnetically dead layer at the LSMO/Nb-STO interface with the thickness $t_D$ of 3.5 $\pm$ 0.6~nm at 2~K. $t_D$ increases with increasing temperature and reaches 6.2 $\pm$ 1.3~nm at 275 K, which is comparable to that in other reports.\cite{Sun1999,Angeloni2004} On the other hand, the $M_\text{s}$ of LSMO/LSAT does not decrease with decreasing LSMO thickness [see Fig.~\ref{fig:Manetization}(b)], suggesting that the dead layer is negligibly thin. It has been reported that lattice strain in LSMO plays an important role in inducing a dead layer,\cite{Liao2019} which is in agreement with our results suggesting the presence of the magnetically dead layer in strained LSMO ($c/a$~=~0.969 -- 0.982) on Nb-STO but not in strain-free LSMO ($c/a\approx$ 1) on LSAT.

$M_{\text{eff}}$ reaches its lowest value near room temperature for all the samples [Figs. \ref{fig:Manetization}(a) and \ref{fig:Manetization}(b)]. For LSMO/Nb-STO, $M_{\text{eff}}$ increases with decreasing temperature to 75~K and shows a sudden drop upon further cooling. Similar to the thickness dependence observed for the damping anomaly, the decrease in $M_{\text{eff}}$ becomes more pronounced with decreasing LSMO thickness and was not observed for LSMO/LSAT without a dead layer. The results suggest that the dead layer near the LSMO/Nb-STO interface is the key to the decrease in $M_{\text{eff}}$. According to the earlier work on LSMO ($x \approx$ 0.3)/STO, the dead layer at the interface becomes antiferromagnetic below 100~K.\cite{Lee2010} The paramagnetic dead layer in our samples can also become antiferromagnetic below 100 K and a magnetic exchange coupling between the antiferromagnetic dead layer and the rest of the ferromagnetic LSMO layer can lead to the effective reduction of $M_{\text{eff}}$. 

A similar behavior has been reported for Ni$_{0.8}$Fe$_{0.2}$/Mn$_{0.5}$Fe$_{0.5}$, in which the presence of an antiferromagnetic layer at the interface decreases $M_{\text{eff}}$ of Ni$_{0.8}$Fe$_{0.2}$,\mbox{\cite{Stoecklein1988}} and it has been pointed out that an antiferromagnetic order along the out-of-plane may be responsible for an increase in the perpendicular magnetic anisotropy field ($H_\text{k}$) and hence the decrease in $M_\text{eff}$ ($M_\text{eff}= M_\text{s} - H_\text{k}/4\pi$).

Since only a small in-plane exchange bias effect has been observed in our sample (see Supplementary Information), the interfacial antiferromagnetic moment may not be fully aligned in-plane but partially ordered along the out-of-plane direction, which can increase $H_k$ and decrease $M_\text{eff}$. Note that $M_\text{s}$ measured by SQUID has not changed at low temperature [see Fig.~6(a)]. Such an interface-induced perpendicular magnetic anisotropy has been also reported in other ferromagnetic/antiferromagnetic heterostructures.\mbox{\cite{Wang2013,Kuswik2015}}

Since the decrease in $M_{\text{eff}}$ occurs at the temperature close to the onset of the anomalous increase of $\alpha$ ($\approx$ 100 K), the damping anomaly can also be related to the antiferromagnetic transition of the dead layer. We attribute the damping anomaly to the spin pumping into the dead layer, which acts as an effective spin sink. The paramagnetic to antiferromagnetic transition of the dead layer reduces its spin diffusion length and mediates an exchange coupling with the rest of the ferromagnetic LSMO, leading to an increased efficiency of the spin pumping and hence to an increase in $\alpha$.

In conclusion, we have demonstrated an epitaxial strain-induced magnetic damping anomaly in LSMO thin films on Nb-STO substrates. The Gilbert damping constant $\alpha$ significantly increases below 100~K along with a decrease in the effective saturation magnetization $M_{\text{eff}}$. This feature is more pronounced for LSMO with a smaller thickness and a larger volume fraction of paramagnetic phase. The results are explained by spin pumping into a magnetically dead layer acting as a spin sink at the LSMO/Nb-STO interface. LSMO on LSAT without a dead layer does not show any signatures of the damping anomaly, suggesting that the strain-induced dead layer plays a key role. The results extend the understanding of the magnetic damping in LSMO thin films and represent a step toward the advancement of energy-efficient spintronic technologies with ultra-low magnetic damping.
\clearpage
\section*{Supplementary Material}
Supplementary material is available for this article. It includes reciprocal space mapping, the exchange bias effect induced by the antiferromagnetic dead layer, and the effect of perpendicular magnetic anisotropy on the reduction of $M_\text{eff}$.

This work was supported in part by JSPS KAKENHI Grant Nos. JP23KK0086, JP21H04614, JP24H00380, JP24K21732, JP20K23374, JSPS International Joint Research Program (JRP-LEAD with UKRI) No. JPJSJRP20241705, JST CREST Grant No. JPMJCR18J1, JST FOREST Grant No. JPMJFR212V, and Iketani Science and Technology Foundation.

\section*{AUTHOR DECLARATIONS}
\subsection*{Conflict of Interest}
The authors have no conflicts to disclose.

\subsection*{Author Contributions}
Ryotaro Arakawa: Data curation (lead); Formal analysis (lead); Investigation (equal); Methodology (equal); Visualization (lead); Writing--original draft (lead). Sachio Komori: Conceptualization (equal); Data curation (supporting); Formal analysis (supporting); Funding acquisition (equal); Investigation (equal); Methodology (equal); Supervision (equal); Visualization (supporting); Writing--review $\&$ editing (equal). Tomoyasu Taniyama: Conceptualization (lead); Funding acquisition (lead); Project administration (lead); Resources (lead); Supervision (lead); Formal analysis (supporting); Investigation (supporting); Methodology (supporting); Writing--review $\&$ editing (equal).

\section*{Data Availability}
The data that support the findings of this study are available from the corresponding author upon reasonable request.

\clearpage
\bibliography{Main_text_revised}

\clearpage

\begin{figure}[h]
    \includegraphics[width=8.5cm]{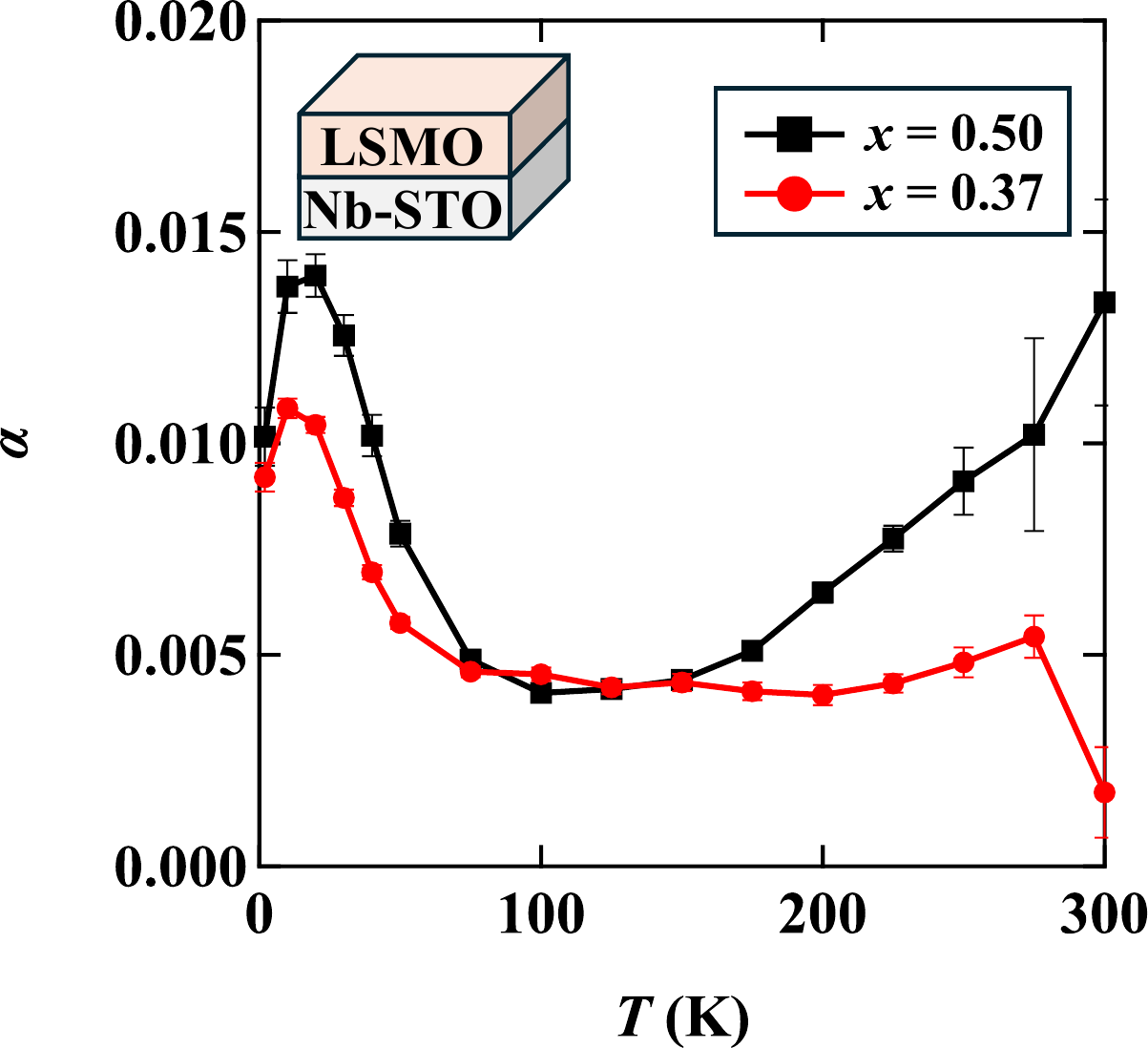}
    \caption{\label{fig:Damping_x} Temperature dependence of the damping constant $\alpha$ for La$_{0.63}$Sr$_{0.37}$MnO$_3$/Nb-STO and La$_{0.5}$Sr$_{0.5}$MnO$_3$/Nb-STO with the film thickness of 38 nm.}
\end{figure}
\clearpage

\begin{figure}[h]
    \includegraphics[width=14cm]{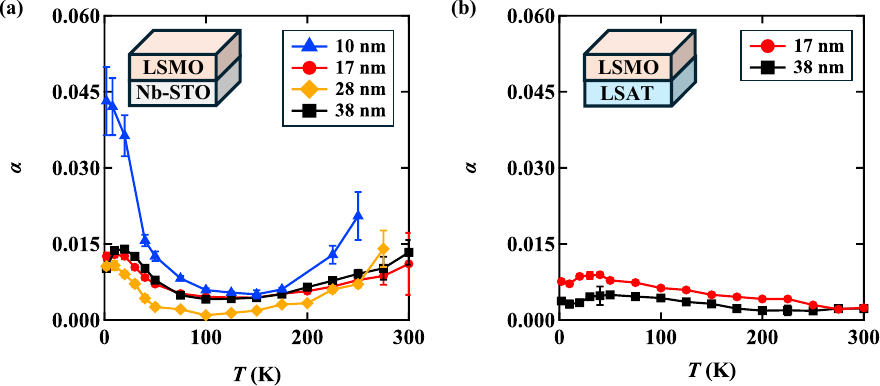}
    \caption{\label{fig:Damping_substrate} Temperature and thickness dependence of the Gilbert damping constant $\alpha$ for (a)~La$_{0.5}$Sr$_{0.5}$MnO$_3$/Nb-STO and (b)~La$_{0.5}$Sr$_{0.5}$MnO$_3$/LSAT.}
\end{figure}

\begin{figure}[h]
    \includegraphics[width=14cm]{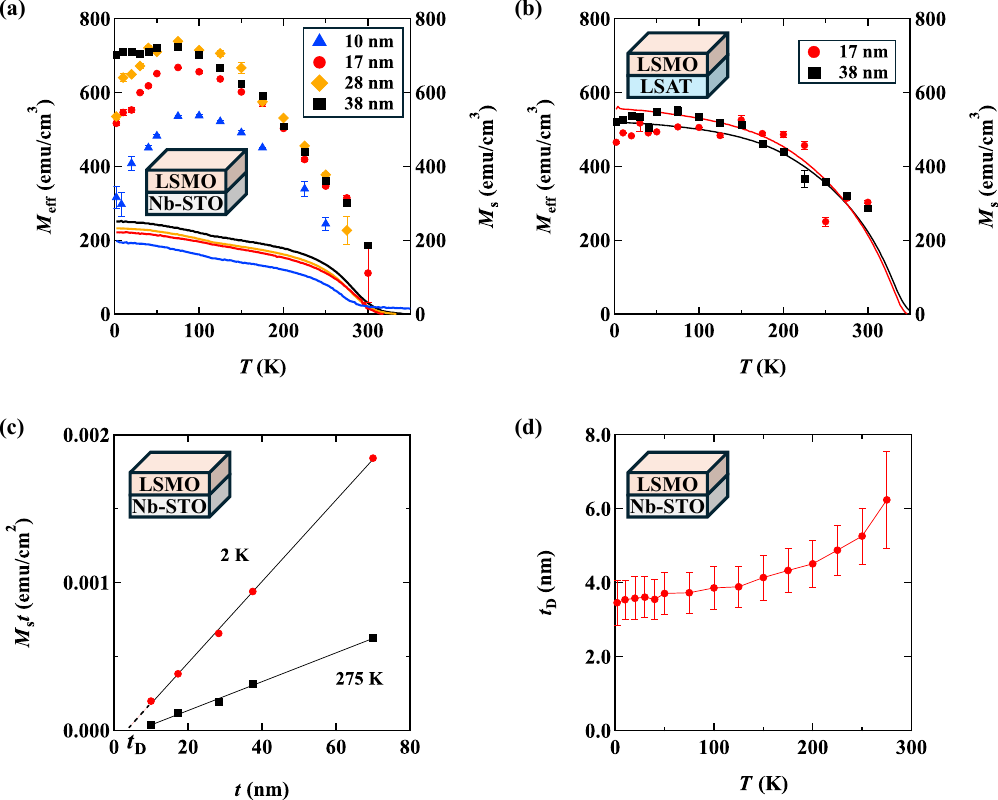}
    \caption{\label{fig:Manetization} Temperature and thickness dependence of the saturation magnetization for (a)~La$_{0.5}$Sr$_{0.5}$MnO$_3$/Nb-STO and (b)~La$_{0.5}$Sr$_{0.5}$MnO$_3$/LSAT. The data points (markers) and the curves represent the effective saturation magnetization $M_{\text{eff}}$ (left-axis) and the DC saturation magnetization $M_\text{s}$ (right-axis), respectively. (c)~Thickness dependence of the saturation magnetization per unit area for La$_{0.5}$Sr$_{0.5}$MnO$_3$/Nb-STO at 2 and 275~K with least-squares line fits, giving a dead layer thickness $t_D$ of 3.5 $\pm$ 0.6~nm and 6.2 $\pm$ 1.3~nm, respectively. (d)~Temperature dependence of $t_D$.}
\end{figure}

\clearpage
\renewcommand{\thefigure}{S\arabic{figure}}  
\setcounter{figure}{0}

\begin{center}
\large \textbf{Supplementary information for\\ Strain-induced magnetic damping anomaly in La$_{1-x}$Sr$_{x}$MnO$_{3}$ ($x$~=~$0.3$~--~$0.5$) thin films}    
\end{center}

\section*{\label{sec:level1}Reciprocal Space Mapping}
\begin{figure}[h]
    \includegraphics[width=12cm]{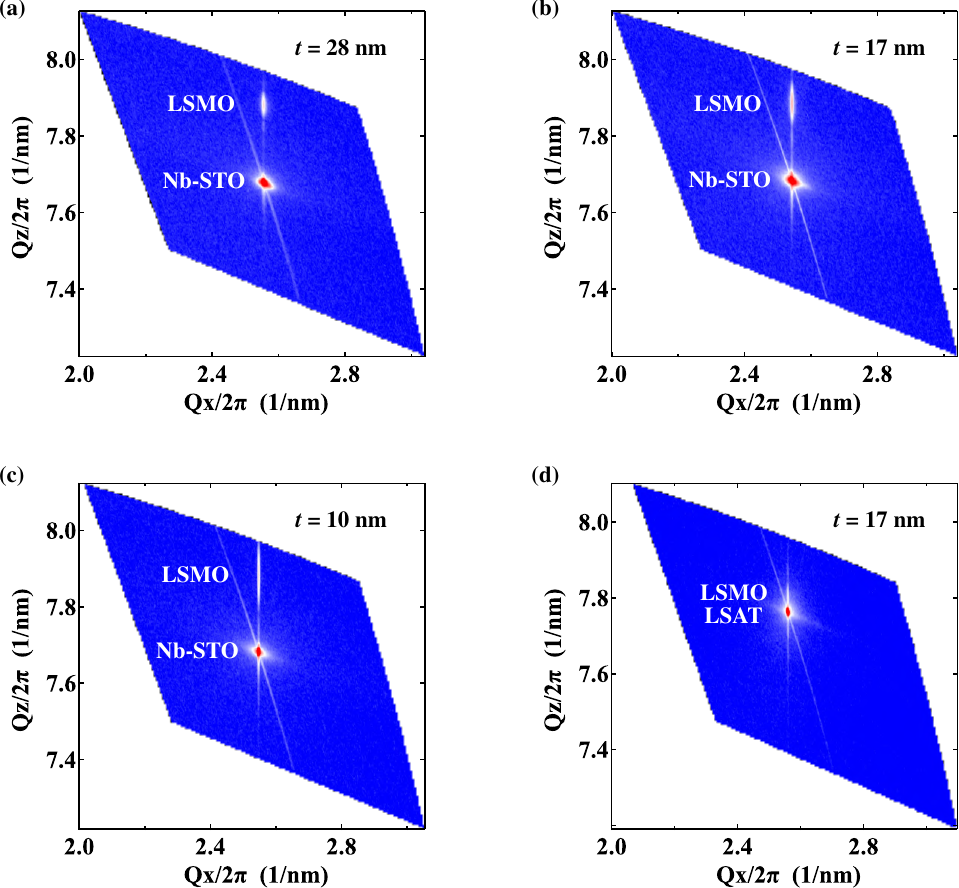}
    \caption{\label{fig:RSM_supplementary} Reciprocal space mapping around the (103) diffraction peak from LSMO ($x$~=~0.5)/Nb-STO with the thickness of (a)~28~nm, (b)~17~nm, (c)~10~nm and (d)~LSMO ($x$~=~0.5)/LSAT with the thickness of 17~nm.}
\end{figure}

Figures \ref{fig:RSM_supplementary}(a--c) show reciprocal space mapping around the (103) peak of 28-nm, 17-nm, and 10-nm-thick LSMO ($x$~=~0.5) on Nb-STO, respectively, confirming coherent growth of LSMO without detectable lattice relaxation in all the samples. 
From the peak position, anisotropic lattice strain ($c/a$ ratio of the lattice constants) of LSMO on Nb-STO is estimated to be 0.972, 0.967, and 0.968, respectively. Figure \ref{fig:RSM_supplementary}(d) shows the data for 17-nm-thick LSMO ($x$~=~0.5) on LSAT, where the peak of LSMO is overlapped with the peak of LSAT, indicating the absence of lattice mismatch.

\clearpage
\section*{Exchange bias effect induced by the antiferromagnetic dead layer}
\begin{figure}[h]
    \includegraphics[width=17cm]{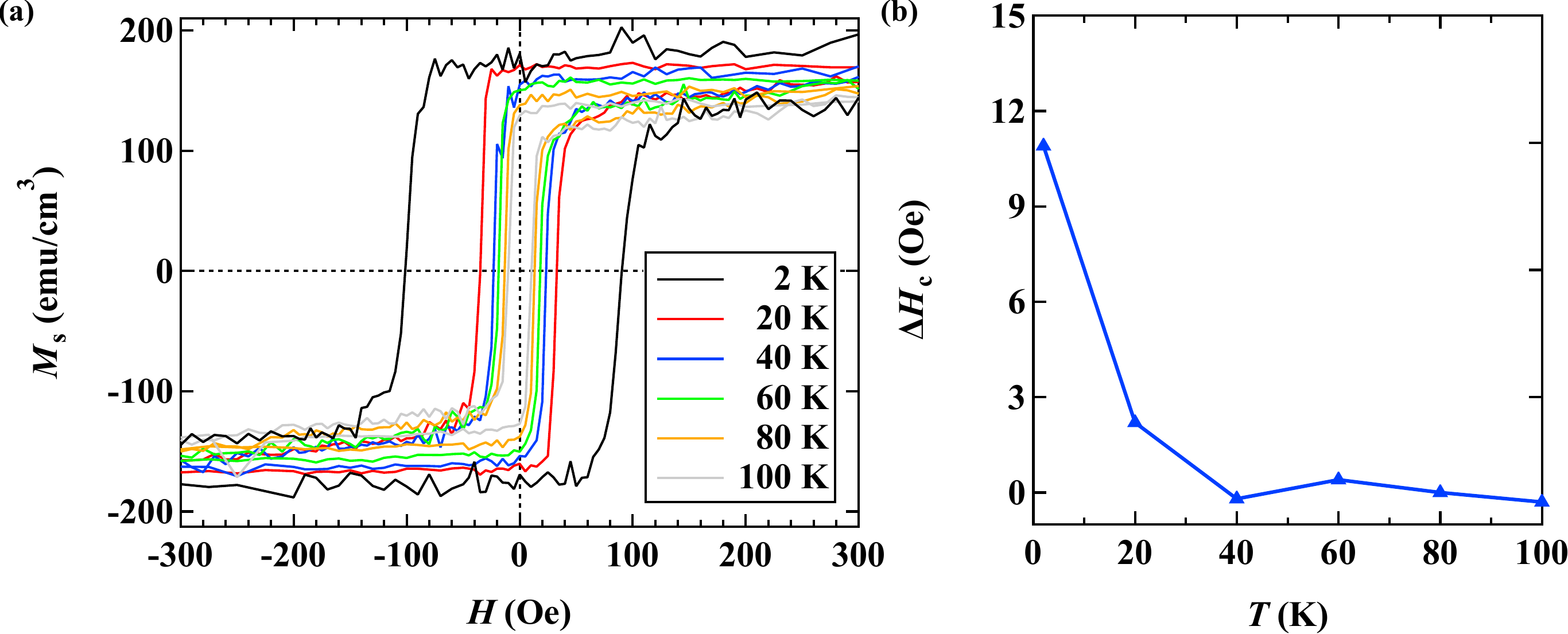}
    \caption{\label{fig:supplementary_MH}  (a) Temperature dependence of the $M$-$H$ curve for 10-nm-thick LSMO ($x$~=~0.5) on Nb-STO measured after field cooling from room temperature at 2~kOe for 20~--~100~K and 20~kOe for 2~K. (b) The difference between positive and negative coercive field ($\Delta H_\text{c}~=~H_\text{c}^{-} - H_\text{c}^{+}$) as a function of the temperature.}
\end{figure}

Figure \ref{fig:supplementary_MH} (a) shows the $M$-$H$ curves measured by SQUID for 10-nm-thick LSMO ($x$~=~0.5)/Nb-STO at 2~--~100~K after field cooling from room temperature.
The difference between the positive and negative coercive field ($\Delta H_\text{c}$~=~$H_\text{c}^- - H_\text{c}^+$) is plotted as a function of temperature (Fig. \ref{fig:supplementary_MH} (b)).
The $M$-$H$ loop is slightly shifted to the left below 40~K, which may be attributed to a small exchange bias effect induced by the antiferromagnetic dead layer.

\end{document}